# Scanning gate imaging of quantum point contacts and the origin of the 0.7 anomaly

Andrea Iagallo[1], Nicola Paradiso[1], Stefano Roddaro[1,2], Christian Reichl[3], Werner Wegscheider[3], Giorgio Biasiol[2], Lucia Sorba[1], Fabio Beltram[1], and Stefan Heun[1] (✉)

[1] *NEST, Istituto Nanoscienze-CNR and Scuola Normale Superiore, Piazza San Silvestro 12, 56127 Pisa, Italy*
[2] *Istituto Officina dei Materiali CNR, Laboratorio TASC, 34149 Trieste, Italy*
[3] *Solid State Physics Laboratory, ETH Zurich, Schafmattstrasse 16, 8093 Zurich, Switzerland*

**ABSTRACT**

The origin of the anomalous transport feature appearing at conductance $G \approx 0.7 \times (2e^2/h)$ in quasi-1D ballistic devices – the so-called 0.7 anomaly – represents a long standing puzzle. Several mechanisms were proposed to explain it, but a general consensus has not been achieved. Proposed explanations are based on quantum interference, Kondo effect, Wigner crystallization, and more. A key open issue is whether point defects that can occur in these low-dimensional devices are the physical cause behind this conductance anomaly. Here we adopt a scanning gate microscopy technique to map individual impurity positions in several quasi-1D constrictions and correlate these with conductance characteristics. Our data demonstrate that the 0.7 anomaly can be observed irrespective of the presence of localized defects, and we conclude that the 0.7 anomaly is a fundamental property of low-dimensional systems.

## 1. Introduction

The study of low-dimensional ballistic systems has yielded a number of exciting observations and has made it possible to investigate several striking physical phenomena during the last 30 years. Despite the conceptual simplicity of the archetypical device, i.e. a quasi-1D constriction (or quantum point contact, QPC), these systems are still attracting much interest both from the point of view of fundamental electron-transport physics and for possible applications in spintronics.

The distinctive feature of 1D ballistic systems is conductance quantization in units of $G_0 \equiv (2e^2/h)$. In the non-interacting picture, these steps are a direct consequence of the progressive population of 1D subbands. The single-particle picture is however unable to explain the 0.7 anomaly, i.e. the occurrence of a plateau-like feature below the last quantized plateau around 0.7 $G_0$ [1, 2]. Though extensively investigated, the origin of this feature remains controversial still today. Several theoretical explanations were put forward over the years, but a universally accepted theory comprising all the experimental features is still lacking (see Ref. [3] for a detailed review). A number of theoretical investigations focused on the role of quantum interference [4, 5], spontaneous spin polarization [1, 6, 7], Wigner crystallization [8] and the Kondo effect [9-11]. Even very recently, two papers provided evidence for the Kondo scenario, based on the formation of a quasi-bound state at the constriction [12, 13]. On the other hand, another recently proposed model involves a "van Hove ridge" in the density of states of the QPC that strongly enhances interaction effects for subopen QPCs without needing spin-polarization or quasi-bound states [14]. A way in which these two scenarios could be consistent despite their apparent differences was suggested by Micolich [15].

On the experimental side, the wealth and variety of observable effects makes a direct comparison between different experiments rather difficult. Moreover, the 0.7 anomaly appears to be rather sensitive to a variety of extrinsic parameters so that its detailed characteristics are typically very much device-dependent [3]. Nevertheless, existing studies so far allowed to clearly spot a set of universal phenomenological properties associated with the 0.7 anomaly which are now universally considered its intrinsic attributes. These include the continuous evolution of the 0.7 structure into a spin-resolved state when a magnetic field is applied, and the strengthening of the 0.7 plateau with increasing temperature [1, 16-18]. Furthermore, experiments strongly suggest a connection between the 0.7 structure and the so-called zero-bias anomaly [19], a peak arising around zero bias in the dI/dV characteristics of QPCs driven near pinch off, suggesting a still debated link between the 0.7 anomaly and Kondo physics [9].

The role and importance of localized impurities and defects in close proximity to the constriction on the physics of the 0.7 anomaly remains rather unclear and is the subject of this article. By localized impurities we intend extrinsic features that have a strong scattering effect in the two-dimensional electron gas (2DEG) and are superimposed on the weak ripples in the background potential caused by remote doping of the 2DEG. Localized impurity is known to strongly perturb channel transmission and can therefore impact 0.7 phenomenology in at least two distinct ways: (i) they could lead to a strongly coupled quantum-(anti)dot potential and drive a Kondo-impurity behavior [20]; (ii) as charge scatterers, they could create a cavity between the constriction and the impurity and thus induce quantum-interference phenomena [21, 22]. Here we investigate the effect of these localized defects (point-defects, impurities) on the 0.7 anomaly.

The subject of localised defects has been dealt with extensively by channel shifting experiments in QPCs [1, 16, 23, 24]. This technique consists in laterally displacing the conductive channel by differentially biasing the metallic gates which define the constriction [25]. Typically, lateral shifts of less than 100 nm are obtained by applying a bias asymmetry below 2 V. If an impurity is present in the region between the gates, its influence on the conductance can be tuned by shifting the channel. This allows to separate the contribution of strong disorder (i.e., backscattering or interference effects) from the conductance of the clean channel. In particular, a plateau appearing at G ≈ 0.7 $G_0$ is considered to be the genuine 0.7 anomaly if, in each and every single trace measured at different value of gate imbalance, it stays at the same conductance value [16]. Any movement to higher/lower conductance or the appearing of new features is judged to stem from impurity [11, 24, 26]. Such a technique is, however, not conclusive since it can detect only defects located at short distance from the constriction. Potential variations occurring either at the side of the constriction or some hundred nanometers outside of it can be invisible to this approach.

In such a configuration, scanning probe microscopy (SPM) can provide additional information. Indeed, SPM techniques are the only methods able to unequivocally detect localized charged structures and their position with respect to the QPC. In the past, a rich sub-$G_0$ spectrum was observed in a QPC fabricated by erasable electrostatic lithography and investigated by scanning gate microscopy (SGM) [27]. The conductance exhibited additional plateaus at ≈ 0.5 $G_0$ and ≈ 0.9 $G_0$, which were attributed to asymmetries in the potential profile of the QPC, probably caused by the unconventional QPC fabrication method. In that study, however, these low-conductance features were not related to the presence of strong disorder.

We make use of SGM and experimentally directly determine whether impurity is present or not near a given constriction. SGM has proven to be extremely sensitive in probing the potential landscape of two-dimensional electron systems [28-30]. This high sensitivity in detecting conductance variations makes the SGM technique an ideal tool to investigate the effect of potential imperfections on the conductance of constrictions. Besides, we exploit the ability of SGM to perform gating [31-33] to carry out enhanced channel shifting measurements where a much larger area is probed, whose radius, for a smooth gate profile, is at least equal to the gate separation width. Indeed, when the biased tip is scanned above the gates around the QPC centre, SGM measurements are equivalent to performing a channel shifting experiment, in that the constriction opening defined by the total potential is moved laterally before pinch-off is reached. In the present work, we thus extend the channel shifting technique to an area of several µm$^2$ around the QPC to locate the position of charged impurities by SGM. Our SGM experimental results allow us to show that the 0.7 structure can be observed in devices without any nearby impurity, and thus we can rule out their role in driving the 0.7 anomaly. Compared to previous SGM investigations of QPCs [27], our data do not show any further plateau except the 0.7 structure, regardless of the strength of the gating and asymmetry induced by the SGM tip.

## 2. Experimental details and methods

Our experiments were performed on 2DEGs obtained from single quantum well GaAs/AlGaAs heterostructures. Schottky split-gate electrodes defining a QPC geometry were patterned by electron beam lithography on top of 2000 µm × 300 µm Hall bars fabricated by standard optical lithography. Ohmic contacts (Ni/AuGe/Ni/Au) and Schottky electrodes (Ti/Au, 10 nm/20 nm) were deposited by thermal evaporation. Experiments were performed with over 10 devices fabricated from different heterostructures in a wide range of electron mobility µ (2 – 12.5 × 10$^6$ cm$^2$/Vs) and density n (1 - 5 × 10$^{11}$ cm$^{-2}$), all consistent with the conclusions presented here. As representative examples, in this work we present data from two devices: $\mu_A$ = 4.64 × 10$^6$ cm$^2$/Vs and $n_A$ = 2.1 × 10$^{11}$ cm$^{-2}$ were measured for device A, while the corresponding values for device B were $\mu_B$ =12.5 × 10$^6$ cm$^2$/Vs and $n_B$ = 2.3 × 10$^{11}$ cm$^{-2}$ (all values measured at 300 mK in the dark). The 2DEG depths $d$ were $d_A$ = 100 nm and $d_B$ = 110 nm, respectively. The split-gate layouts used were nominally identical in shape for all devices, differing only in the constriction width $w$, which were $w_A$ = 500 nm and $w_B$ = 400 nm, respectively.

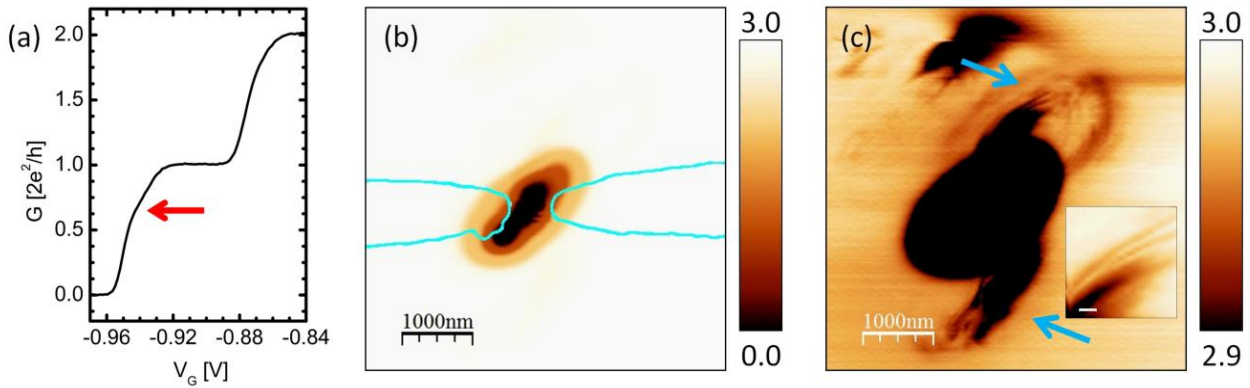

**Figure 1** Device A: QPC with localized impurities. (a) Conductance of device A as a function of gate voltage $V_g$, at T = 300 mK. The 0.7 anomaly appears as a shoulder below the last two conductance plateaus (see arrow). (b) SGM scan of the differential conductance G of the QPC area of device A. T = 300 mK, $V_g$ = -0.78 V, $V_{tip}$ = -2 V, and QPC conductance G = 3 $G_0$ (3rd plateau). The three lowest conductance plateaus are visible as concentric structures. The outline of the split-gates as obtained from an AFM topography image is indicated by the blue lines. (c) Same data as in (b), after optimizing the contrast to highlight the presence of impurities (lower arrow) and of additional features associated with (anti)dot formation (upper arrow). For both images, G is given in units of $G_0 \approx (2e^2/h)$. Inset: an enlarged view of the area indicated by the upper arrow shows Coulomb blockade oscillations of G with HWHM = 25 nm. The scale bar is 100 nm long, the conductance range shown is (2.8 – 3.0) $G_0$.

SGM experiments were performed by scanning the metallic tip of an atomic force microscope (AFM) over the surface of the devices at fixed height ($d_{tip} \approx 40$ nm), sufficient to ensure a distance of at least 10 nm from the metallic split-gates. Local gating was induced by applying a suitable negative voltage $V_{tip}$ to the AFM tip, which created a potential perturbation of ~150 nm half-width at half-maximum (HWHM). For the experiments, we used tungsten tips electrochemically etched in a 2 M NaOH solution with homebuilt electronics. The measurements were carried out in a $^3$He cryostat. The QPCs were first localized by acquiring the topography of the devices. Then SGM measurements were performed by recording the source-drain conductance of the Hall bar as a function of tip position. The conductance was measured by standard lock-in technique, and the data were processed with the aid of the WSxM software [34].

In SGM investigations of 2DEG constrictions the tip perturbation is often larger than the size of the conductive channel. Despite this fact, a resolution down to the Fermi wavelength is typically achieved because the lateral resolution of these experiments is not determined by the radius of the tip perturbation but by the precision in positioning it [28]. This is understandable when the tip is at large distance from the QPC centre, and the tip- and gate-induced potentials are independent. In such a case, the resolution in detecting variations in the background potential is limited by the accuracy in setting the distance between two scatterers (e.g., the tip and an impurity). Such a fine resolution was also demonstrated to be ~ 10 nm in our setup by visualizing coherent interference fringes in branched electron flow [35] and fractional incompressible stripes in the quantum Hall regime [33]. On the other hand, in proximity to the QPC centre, the 2DEG is subject to the total potential given by the sum of the fixed gates and the tip. Nevertheless, in such a condition, the minimum resolution of the measurement is set by the accuracy in shifting the position of the conducting channel, which is ultimately set by the accuracy in positioning the charged tip. Screening of the tip potential by the metallic gates plays a minor role in our experiments, because the 2DEG responds to the total potential at the constriction. Not surprisingly, as we show later, we observe conductance variations also when the tip centre is above one of the gates. We can exclude that this effect is the result of galvanic contact between the tip and the gates, since such an occurrence would cause irreversible damage of the device.

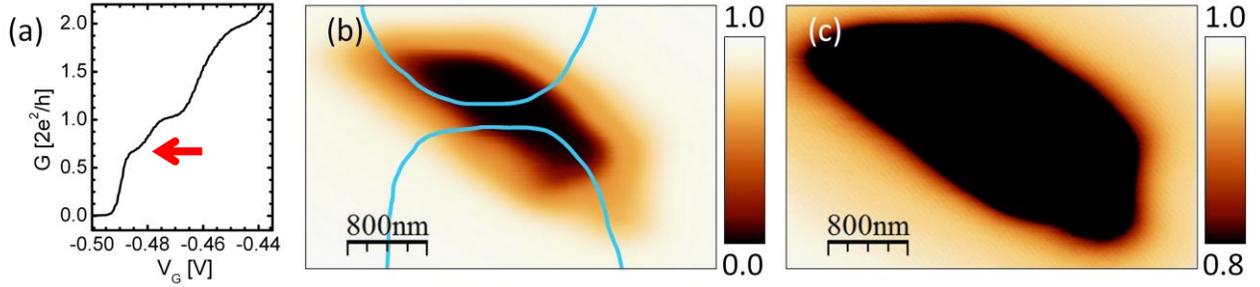

**Figure 2** Device B: QPC without localized impurities. (a) Source-Drain conductance of device B as a function of gate voltage, displaying a clear 0.7 anomaly at $G \approx 0.65\ G_0$. (b) SGM scan of the differential conductance G of the QPC area of device B. $T = 1$ K, $V_g = -0.47$ V, $V_{tip} = -4.25$ V, and QPC conductance $G = G_0$ (1st plateau). The outline of the split-gates, obtained from an AFM topography scan, is indicated by the blue lines. (c) Same data as in (b), after optimizing the contrast, showing a QPC constriction without any localized defect. For both images, G is given in units of $G_0 = (2e^2/h)$.

## 3. Results and discussion

In this work, we first consider the presence of localized impurities and their effect on the sub-$G_0$ conductance. In Fig. 1(a), the source-drain conductance through the QPC of device A is shown as a function of gate voltage $V_g$. Below the last conductance plateau, the 0.7 anomaly is observed as a shoulder appearing at $G \approx 0.7\ G_0$ (see arrow). Figure 1(b) shows an SGM image of device A obtained by scanning the QPC area with a voltage $V_{tip} = -2$ V applied to the tip. For the measurements, the gates were biased with a voltage $V_g = -0.78$ V. The image clearly displays the lowest three conductance plateaus, visible as annular concentric structures, and QPC pinch-off. We have observed coherent branched electron flow on similar samples [33]; however, it is not observed in the images presented in this article because the applied tip voltages did not completely deplete the electron gas underneath the tip [31]. The asymmetry of the pinch-off spot in Fig. 1(b) reflects a small asymmetry of the tip potential, caused by a not perfectly circular section of the tip possibly resulting from the etching process. However, as long as the tip perturbation has smooth equipotential lines, this aspect is of no relevance in the SGM measurements discussed here.

Localized defects close to the QPC center can be spotted in high-contrast SGM maps, as shown in Fig. 1(c). The lower arrow in Fig. 1(c) shows one impurity near the QPC entrance that appears in the SGM image as a dark spot and identifies an area where a sharp potential variation occurs. The upper arrow in Fig. 1(c) shows an additional structure: concentric ring-like shapes indicate the formation of an (anti)dot [30, 36-41] owing to a modulation of either quantum-interference or Coulomb-blockade effects by the SGM tip. The ability to detect this modulation, better seen in the inset in Fig. 1(c), demonstrates a lateral resolution of the microscope better than the HWHM of the structures, 25 nm. Taken together, the results of Fig. 1 demonstrate two important points: 1) when impurities are present, their signatures are indeed detected thanks to the high sensitivity of the SGM technique; 2) the 0.7 anomaly is observed in the presence of impurities located within short distance from the QPC. As such, our results show that the 0.7 anomaly is quite robust to strong perturbations of the saddle point potential of the QPC, though they do not imply that impurities are *necessary* for its observation. Indeed, as we show in the following, a very clear effect can also be observed in the case of sample B, where no impurity is present.

The QPC conductance of device B is shown in Fig. 2(a) as a function of $V_g$. In this device, the 0.7 anomaly manifests as a marked shoulder appearing at G ≈ 0.65 $G_0$ (see arrow) in an otherwise smooth conductance trace. Figure 2(b) shows an SGM image of the QPC area of device B, where the conductance ranges from complete pinch-off to the first quantized plateau at G = $G_0$. Beyond the standard SGM-induced electrostatic depletion of the constriction, we detect no trace of sharp potential fluctuations: differently from the case of sample A, we can rule out the presence of localized impurities in the scanned area. This becomes even more evident when increasing the contrast of the SGM image, as displayed in Fig. 2(c).

In order to completely exclude interference effects between a scatterer and the QPC, we carefully scanned an even larger area around the QPC. It is well known that only those impurities within half a thermal length $\ell_{th}$ of the QPC are able to interfere with the QPC [32, 42, 43]. Interference effects can in fact occur at distances from the QPC much longer than the thermal length: for instance, this is the case of two impurities separated by a distance smaller than $\ell_{th}$, but located at a distance >> $\ell_{th}$ from the QPC. This situation is very different from the one considered in the present work, because, although the electrons backscattered from the two impurities can interfere between them, they lose coherence well before reaching the QPC area, i.e. cannot interfere with the QPC, and thus have no effect on the 0.7 anomaly. In our experiment, the thermal length $\ell_{th}$ = $2\pi\hbar^2/(m\lambda_F k_B T)$ ≈ 4 μm, where m and $\lambda_F$ are the effective mass and Fermi wavelength of the electrons in the 2DEG, respectively. We thus systematically checked an area of more than 5 μm radius around the QPC centre, and no localized scatterers were detected.

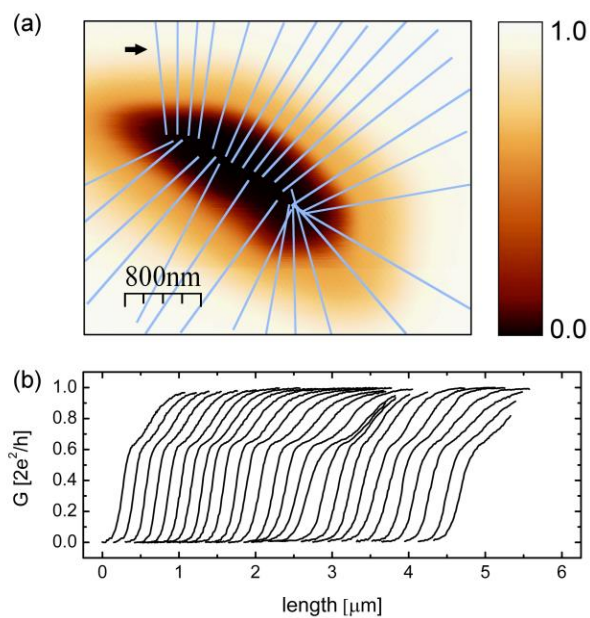

**Figure 3** 0.7 anomaly in device B. (a) SGM scan of device B measured for T = 0.7 K, $V_g$ = -0.471 V, $V_{tip}$ = -5 V, showing the low conductance range. The conductance is given in units of $G_0$ =($2e^2/h$). (b) Conductance cross-sections obtained from the SGM scan in (a), offset for clarity. The traces, from left to right, are cuts along the lines in (a), starting at the arrow and following the clockwise direction. The 0.7 anomaly is at the same conductance value G = 0.63 $G_0$ in all traces.

Since in device B the 0.7 anomaly is so pronounced, it can be identified directly from an SGM map. Figure 3(b) shows several conductance cross-sections cut along the lines highlighted in Fig. 3(a). These lines were chosen to be perpendicular to the conductance edges. All profiles display a shoulder at G = 0.63 $G_0$ and no further features, consistently with the results of the transport measurements in Fig. 2(a). In each trace, this shoulder appears within a narrow range ΔG = ± 0.0015 $G_0$ around the value G = 0.63 $G_0$. This very small range of conductance variations is comparable to the typical variations of the 0.7 anomaly obtained in transport measurements performed on clean constrictions, and consistently with the channel shifting measurements found in literature [16, 24, 26] it clearly demonstrates that no impurity is present at short distance from the channel. A smaller steepness of the risers between plateaus is observed at certain directions, probably as a result of a smoother confining potential. However, also for these curves, the sub-$G_0$ structure appears at the same conductance value G = 0.63 $G_0$. Furthermore, we note that the 0.7 anomaly retains the same value regardless of the direction, thus appearing to be an isotropic property of our device. We emphasize that by employing the SGM technique, we perform a 2D channel shifting experiment which allows to probe an area of the order of 1 $\mu m^2$, against the typical ~ 100 nm 1D shift of traditional channel shifting measurements. This gave us the possibility to observe the robustness of the 0.7 anomaly at different directions around the pinch-off region, an aspect which cannot be tested by traditional transport measurements.

The fact that we clearly observe the 0.7 anomaly in a constriction free of defects, both in transport and in SGM measurements of the same device, is the main finding of this investigation. We can therefore discard any impurity-related mechanisms as the origin of the 0.7 anomaly.

The present analysis reduces the number of possible candidate mechanisms for the 0.7 anomaly and demonstrates that an explanation of the phenomenological attributes (e.g., Kondo physics and spin polarization) must relate to intrinsic properties of the constriction. The formation of a self-consistent state inside the constriction hosting a localized spin was proposed [9, 10, 44] to explain the evolution of the 0.7 structure with magnetic field. In this scenario, at low temperature the resulting unpaired spin is screened by the conduction electrons due to Kondo physics. At high temperatures, the screening becomes less effective, and produces a characteristic value of conductance 0.5 $G_0$ < G < 1 $G_0$. This localized state would provide a tool to perform spin manipulation by electrostatic means, giving a significant contribution to the development of nonmagnetic spintronic and spin-selective devices [45-47].

Our SGM investigation on impurity-free QPC constrictions opens the way to further scanning probe microscopy experiments dedicated to point the way towards other possible theoretical origins of the 0.7 anomaly. One could use the SGM technique to change the strength of the 0.7 anomaly, using the tip to accurately spatially control the potential profile within the QPC. For instance, in order to verify the existence of this quasi-bound state inside the constriction, the tip could be used to selectively populate/depopulate the proposed quasi-bound state. Very recently, SGM experiments pointing in this direction were reported on quasi-1D constrictions [13] where fluctuations in the conductance maps were interpreted as the signature of a length-dependent chain of spin-polarized states forming within the constriction channel. These results are compatible with ours and strengthen the link between the 0.7 structure and the dimensionality of quasi-1D devices, and highlight the potential of SGM for the investigation of electron interactions in these systems.

Another route could be an SGM investigation of the effects of the barrier geometry on the 0.7-anomaly. Such an experiment could effectively test the scenario involving a "van Hove ridge" in the density of states of a QPC [14]. In this model, the sample geometry is extremely important for the precise way in which the density of states affects the conductance.

## 4. Conclusions

In summary, we studied the occurrence of the 0.7 anomaly in QPCs with and without impurity-related localized potential fluctuations identified by SGM imaging. We observed the 0.7 structure with and without charged defects in proximity to the constriction and showed that it presents annular symmetry around the depleted spot at the QPC centre. These experiments show that any physical models based on localized defects (i.e. interference effects and Kondo effect due to localized quantum (anti)dots) for the 0.7 structure are not correct and that the latter is an intrinsic property of low-dimensional systems.

## 5. Acknowledgements

This work was supported by the Italian Ministry of Research (MIUR-FIRB project RBID08B3FM) and by the Italian Ministry of Foreign Affairs (Ministero degli Affari Esteri, Direzione Generale per la Promozione del Sistema Paese, progetto: Nanoelettronica quantistica per le tecnologie delle informazioni).